\documentstyle[preprint,aps]{revtex}
\begin{document}
\draft
\begin{title}
{
 Theory of tunnelling into and from cuprates }
\end{title}
\author{A.S. Alexandrov}
\address
{Department of Physics, Loughborough University, Loughborough LE11
3TU, U.K.}

\maketitle
\begin{abstract}
A single-particle spectral density is proposed for cuprates taking into
account the  bipolaron formation,
realistic band structure, thermal fluctuations and disorder.
Tunnelling  and photoemission (PES) spectra  are described, including
the temperature independent gap observed both in the superconducting and
normal states, the emission/injection asymmetry, the finite zero-bias
conductance, the spectral shape in the gap region and its temperature
and doping dependence,  dip-hump incoherent asymmetric features at
high voltage (tunnelling) and large binding energy (PES).
\end{abstract}
\narrowtext

The strong-coupling extension of the BCS theory based on the $1/\lambda$
multi-polaron perturbation technique  firmly predicts the
transition to a charged Bose liquid in the crossover region of
intermediate values of
  the BCS
coupling constant $\lambda$ \cite{ale}.
There is  a  fundamental difference between the bipolaron theory of
high-$T_{c}$ cuprates \cite{ale1} and  other theories involving
   real-space pairs (bosons)
  tightly bound by a field of a pure electronic origin. As emphasised by
  Emery $et$ $al$ \cite{em} such `electronic' theories
  are $a$ $priori$ implausible due
  to the strong short-range Coulomb  repulsion between two carriers. The
  direct (density-density) repulsion  is usually
   much stronger than any exchange interaction.
     On the other hand, the $Frohlich$
 electron-phonon interaction can provide
 $mobile$ intersite  bipolarons in the
    $CuO_{2}$ plane condensing  at  high $T_{c}$  \cite{ale2}.
    The  (bi)polaronic nature of carriers in  cuprates explains a very
    small coherence volume in the superconducting state,
 the mid-infrared  conductivity \cite{mul},
 the  isotope effect on the carrier mass\cite{mul2}.

 Although   the charged Bose liquid of bipolarons  describes
 anomalous thermodynamics and  kinetics
 of  cuprates \cite{alemot},
finite frequency/momentum response functions
  of bipolaronic superconductors remain  to be established. In this letter
  we derive a single-particle spectral function of strongly coupled
  carriers in a random potential which
  provides
  a quantitative description of recent tunnelling spectra\cite{ren,dew}
  and explains some
  photoemission features (see \cite{she} and references in
  \cite{ren,dew,she}).

In the framework of our theory the ground state of cuprates is a charged
Bose-liquid of intersite bipolarons with single polarons existing only as
excitations with the energy $\Delta/2$ or larger. Different from the BCS
description the pair binding energy $\Delta$ is temperature independent.
Hence, there is no other phase transition except a superfluid one at
$T=T_{c}$. The characteristic temperature $T^{*}$ of the normal phase is a
crossover temperature of the order of $\Delta/2$ where the population of
the upper polaronic band becomes comparable with the bipolaron density.
Along this line the theory of tunnelling in the bipolaronic
superconductors was developed both for two-particle \cite{aletun} and
single- particle \cite{rub} transitions through a dielectric contact.
It allowed us to understand the temperature independent gap and the
asymmetry of the
 current-voltage characteristics,  observed already
 in the earlier tunnelling experiments \cite{gee}. However, an attempt to
fit the
conductance  structure  led us to a very narrow (bi)polaronic
band with a bandwidth of the order or even less than $T_{c}$ \cite{aletun,rub}.
Such a bandwidth is not
compatible  with the experimental estimate of the effective
carrier mass, $m^{*} \simeq 2 - 10 m_{e}$ (depending on doping) from the London
penetration depth \cite{alemot}). It is also incompatible with the
theoretical estimate
\cite{ale2} of the (bi)polaron bandwidth (c.a. $100 meV$ or larger)
based on  the well-established
value of the Frohlich
interaction. Moreover, any description based on the Bloch representation
is hardly justifiable for cuprates with the mean free path often
comparable
with the lattice constant.
  One has to consider
 a random potential and thermal fluctuations
 along with a strong pairing potential and
 band-structure effects.

 We apply a single-particle tunnelling Hamiltonian describing the
 injection  of an electron into a single hole polaronic state $P$ with the
 matrix element $P_{{\bf k},\nu}$ and into a paired hole (bipolaronic) state
$B$ with the matrix element $B_{{\bf k},\nu,\mu}$  (Fig.1),
\begin{equation}
H_{tun}=\sum_{{\bf k},\nu}P_{{\bf k},\nu} a_{\bf k}p_{\nu} +\sum_{{\bf
k},\nu,\mu}B_{{\bf k},\nu,\mu}a_{\bf k}p^{\dagger}_{\nu}b_{\mu} +H.c.
\end{equation}
Here ${\bf k}$, $\nu$ and $\mu$ are the quantum numbers describing the
electron in a tip (Fig.1) (annihilation operator $a_{\bf k}$)
a hole polaron
 ($p_{\nu}$) and a hole bipolaron ($b_{\mu}$) in the
$CuO_{2}$ plane in a
random field, respectively. If the eigenstates
$ |\nu \rangle$ and $|\mu \rangle$ are known,
the matrix elements $P_{{\bf k},\nu}$ and
$B_{{\bf k},\nu,\mu}$ are derived by the use of the site
representation and the canonical polaronic transformation   as
 discussed in detail in Ref. \cite{aletun,rub}.
 They are almost independent of $k\simeq k_{F}, \mu, \nu$
in a wide
voltage and binding energy range, $P_{{\bf k},\nu}\simeq P=constant$,
 $N^{1/2}B_{{\bf k},\nu,\mu}\simeq B=constant$, $N$ is the number of cells
 in the sample volume. In general, $B$ and
 $P$ are different \cite{rub} because  the second hole
in a small coherence volume changes the potential  barrier of the contact
for the tunnelling $B$ compared with $P$.

The injection rate is given by the Fermi Golden Rule as
\begin{eqnarray}
W_{in}& =& 2\pi P^2 \sum_{{\bf k},\nu} f(\xi_{\bf k})f(\xi_{\nu})
\delta (\xi_{\bf k}+eV+\xi_{\nu}+\Delta/2)\cr
&+& {2\pi B^2\over{ N}} \sum_{{\bf k},\nu,\mu}f(\xi_{\bf
k})n_{\mu}[1-f(\xi_{\nu})]\delta (\xi_{\bf
k}+eV+\xi_{\mu}-\xi_{\nu}-\Delta/2),
\end{eqnarray}
and the emission rate is
\begin{eqnarray}
W_{em}& =& 2\pi P^2 \sum_{{\bf k},\nu} [1-f(\xi_{\bf
k})][1-f(\xi_{\nu})]
\delta (\xi_{\bf k}+eV+\xi_{\nu}+\Delta/2)\cr
&+&{ 2\pi B^2\over{N}} \sum_{{\bf k},\nu,\mu}[1-f(\xi_{\bf
k})][n_{\mu}+1]f(\xi_{\nu})\delta (\xi_{\bf
k}+eV+\xi_{\mu}-\xi_{\nu}-\Delta/2).
\end{eqnarray}
Here $f(\xi)=[exp(\xi/T)+1]^{-1}$ is the Fermi-Dirac distribution
function for electrons in the tip and hole polarons in the sample,
$n_{\mu}$ is the bipolaron distribution \cite{ref}, $\xi_{{\bf
k},\nu,\mu}$ is the energy spectrum of electrons, polarons and
bipolarons, respectively, $V$ is the sample voltage, $e$ is the
elementary (positive) charge, and $\hbar=k_{B}=1$. Taking the difference
of the injection and emission rates multiplied by $e$ and differentiating
it
with respect to $V$  we obtain the conductance $\sigma =dI/dV$ as
\begin{eqnarray}
\sigma(V,T)&=&{\pi
N_{tip}P^{2}\over{T}}\int_{-\infty}^{\infty}d\xi
\rho(\xi)sech^{2}\left[{\xi+eV+\Delta/2\over{2T}}\right]\cr
&+&{\pi
N_{{tip}}B^{{2}}\over{TN}}\int_{-\infty}^{\infty}d\xi
\rho(\xi)\sum_{\mu}[n_{\mu}+f(\xi)]sech^{2}\left[{\xi-\xi_{\mu}-eV+\Delta/2\
over{2T}}\right],
\end{eqnarray}
with $N_{tip}$ and $\rho(\xi)$ the density of states (DOS) per spin in the tip
near the Fermi level and the polaronic DOS in the sample,
respectively. To arrive at the analytical result we assume that
 the Coulomb bipolaron-bipolaron repulsion\cite{alemot} is relatively
 weak, and temperature is small compared with $\Delta$ and the characteristic
 energy scale $\epsilon_{0}$ of the polaronic DOS (see below). Then  the
bipolaron distribution $n_{\mu}$ is
 narrow \cite{ref} and the population of the polaronic states is low.
 That allows us to integrate out the bipolaronic states  taking
$\xi_{\mu}=f(\xi)=0$ in the second
 term of Eq.(4). The derivative of the Fermi distribution,
 $sech^{2}(x)$ cuts the integrals at $|x|=1$.
 As a result we obtain
 \begin{eqnarray}
 \sigma(V,T)&=& {\pi e^2 N_{tip}P^{2}\over{T}}
 \left[ N (-eV-\Delta/2+2T)-N (-eV-\Delta/2-2T)\right]\cr
&+&{\pi e^2 x N_{tip}B^{2}\over{T}} \left[N
(eV-\Delta/2+2T)-N(eV-\Delta/2-2T)\right],
 \end{eqnarray}
with $N(E)= \int_{-\infty}^{E} dE'\rho(E')$ the cumulative polaronic DOS,
and $x$  the bipolaron density per cell proportional to the doping.
The first term in Eq.(5) proportional to $P^2$
 contributes to the emission, while the second one describes the injection.
They have different values because, in general, we have $B \geq P$ and
$x\leq 1$, so that
 $A\equiv P^2/B^2x$ differs from unity
\cite{rub}. We notice  that the spectral shape of the emission and the
injection might be also different if the bipolaron distribution is
sufficiently wide. There is an additional
integration in the second term in Eq.(4), describing the injection
 compared with the emission.
Hence, a fine
spectral structure like the dip found in the emission  \cite{she,dew} (and
references therein)
can be nearly washed out from the injection as observed \cite{ren,dew}.

The tunnelling (or PES) spectrum, Eq.(5) is defined if the
polaronic cumulative DOS is known. It depends on the band structure,
dressing and scattering. While high frequency phonons and magnetic
fluctuations are
responsible for the  high-energy spectral features in the region of the
order of  the Franck-Condon (polaronic) level shift ($> 100$ meV)
\cite{alemot},
the low energy spectral function is determined by the band
structure and by the thermal lattice, spin and random  fluctuations.
 The $p$-hole polaron in cuprates is almost
one-dimensional due to the large difference in the $pp\sigma$ and $pp\pi$
hopping integrals and the effective `one-dimensional'
localisation  by the random
potential as described in Ref.\cite{ale2}.
This  is  confirmed by the
angle-resolved PES \cite{cam} with no dispersion along certain
directions of the two-dimensional Brillouin zone. Because the amount of
disorder is high  and the screening radius is about  the
lattice constant, we can describe the effect of disorder and of the
thermal  fluctuations as `white Gaussian noise'.  The
relevant  spectral density $A(k,E)$ for a one-dimensional particle in
a random Gaussian potential
was derived by Halperin \cite{hal} and the density of states, $\rho(E)$
 by Frish and
Lloyd \cite{fri}. Halperin found the spectral function both for a
`Schrodinger' particle (i.e. in the effective mass approximation) and for a
`discrete' particle (tight-binding approximation). The estimated polaronic
bandwidth is about $100$ meV or larger, so we consider the effective mass
($m^{*}$) approximation for the spectral density, given by \cite{hal}
\begin{equation}
A(k,E)= 4\int_{-\infty}^{\infty}p_{0}(-z)Re p_{1}(z)dz,
\end{equation}
with $p_{0,1}(z)$ obeying  two differential equations
\begin{equation}
\left[{d^2\over{dz^2}}+{d\over{dz}}(z^2+2E)\right]p_{0}=0,
\end{equation}
and
\begin{equation}
\left[{d^2\over{dz^2}}+{d\over{dz}}(z^2+2E)-z-ik\right]p_{1}+p_{0}=0.
\end{equation}
The boundary conditions are to be found in Ref. \cite{hal}. Here the energy $E$
and the momentum $k$ are measured in the units
$\epsilon_{0}=(D^{2}m^{*})^{1/3}$ and $k_{0}=(D^{1/2}m^{*})^{2/3}$,
respectively.
The constant $D$ describes the second moment of the Gaussian potential
comprising  thermal and random fluctuations as
$D=2(V_{0}^{2}T/M+n_{im}v_{0}^{2})$, where $V_{0}$ is the amplitude of
the deformation  potential, $M$ is the elastic modulus, $n_{im}$ is the
impurity density, and $v_{0}$ is the coefficient of the $\delta$-
function impurity potential.
Then the cumulative DOS
\begin{equation}
N(E)=(2\pi)^{-1}\int_{-\infty}^{E}dE'\int_{-\infty}^{\infty}dk A(k,E')
\end{equation}
is expressed analytically in terms of the tabulated Airy
functions $Ai(x)$ and $Bi(x)$ as
\begin{equation}
N(E) = \pi^{-2}\left[Ai^{2}(-2E) +Bi^{2}(-2E)\right]^{-1}.
\end{equation}
Substituting Eq (10) into Eq.(5) we obtain the  expression for the
tunnelling and PES spectra in the  voltage (energy) region, where the
high-frequency phonon
 shake-off is forbidden by the energy conservation. In
particular, we find for $T=0$
\begin{equation}
{\sigma(V,0)\over{\sigma_{0}}}=  A
\rho\left({eV-\Delta/2\over{\epsilon_{0}}}\right)+
\rho\left({-eV-\Delta/2\over{\epsilon_{0}}}\right),
\end{equation}
with
\begin{equation}
\rho(x)={4\over{\pi^{2}}}\times{Ai(-2x)
Ai'(-2x)+
Bi(-2x)Bi'(-2x)\over{
[Ai^{2}(-2x)
+Bi^{2}(-2x)]^{2}*}}
\end{equation}
and the constant
$\sigma_{0}=2\pi k_{0}\epsilon_{0}^{-1}e^2 x N_{tip}B^{2}$.

We compare the  conductance, Eq.(11) with the scanning tunnelling
microscope (STM) \cite{ren} and point-contact tunnelling (PCT) \cite{dew}
measurements in an overdoped and optimally doped
$Bi_{2}Sr_{2}CaCu_{2}O_{8+\delta}$ (Bi-2212) in Fig.2 and Fig.3, respectively.
The bipolaron theory  describes quantitatively the spectra in
the gap region, including the zero-bias conductance at $T=0$, the asymmetry, and
the decreasing background at higher voltages which are inconsistent with
 the classic
BCS theory, no matter $s$ or $d$-wave.  The zero-bias conductance at
$T=0$ is explained by
the presence of the impurity tails of the polaronic DOS inside the gap,
while the decreasing background, proportional to $|V|^{-1/2}$ for $|V|
>>\Delta/2$ is explained by the one-dimensional band dispersion of
polarons. The peak amplitudes and the
zero-bias conductance are  determined by
the ratio of the bipolaron binding energy $\Delta$ to the characteristic
scattering rate $\epsilon_{0}$. The  position of two peaks and their
amplitudes
relative to the zero-bias
conductance allow us to determine the relevant parameters $\Delta,
\epsilon_{0}$ and the asymmetry $A$ with an error bar less than
five percent.
The doping
dependence of $\Delta$ agrees with that found  from  the
uniform magnetic susceptibility and
explained by us \cite{alekab} as  resulting from
the screening of the Frohlich interaction by free carriers.

An essential feature of the
bipolaron theory is the temperature independent gap with the ratio
 $\Delta/T_{c}$, which might be quite different from the BCS one,
 $2\Delta_{BCS}/T_{c}\simeq 3-5$.
We  obtain a very large ratio, $\Delta/T_{c}\simeq 9$
 for both overdoped, Fig.2 and
 optimally doped, Fig.3, samples. Such a large ratio
  is difficult to understand in the framework of
  the BCS theory including its
 canonical strong-coupling extension.
 Renner $et$ $al$\cite{ren} emphasised that the evolution of the STM
 spectra with
 temperature was  very different from the classic BCS
 behaviour, both $s$ and $d$-wave. The
 `superconducting' gap was found temperature independent evolving into
the `normal' gap above $T_{c}$. In our theory this is one and the same
gap, which is the bound energy of real-space bipolarons.  It does not
disappear at any temperature, neither at $T_{c}$ nor at $T^{*}$. The
theoretical evolution of the spectrum with temperature described in
Eq.(5),
Fig.3 (inset), reproduces
well the  experimental features. In particular, the zero-bias
conductance increases with temperature and there is no sign that the gap
closes at a given temperature, in agreement with the experiment\cite{ren}. The
latter observation rules out any role of superconducting phase or spin
fluctuations in the normal gap. We notice that the theoretical
peaks shift to higher energies above $T_{c}$, Fig.3 (inset), as observed
(see Fig.2 in Ref. \cite{ren}).

There is some characteristic voltage (binding energy) $V_{c}$, (Fig.2 and
Fig.3), above which the experimental STM and PCT conductance  deviate
from the theoretical one. We believe that a hump  observed
above $V_{c}$ is due to a polaronic cloud, as discussed
 in Ref. \cite{alemot}. The high-frequency phonons and magnetic
fluctuations contribute to the excess spectral weight with a maximum around
 twice  the Franck-Condon shift. The dip
 of the spectral weight at $V_{c}$ is explained by the
electron-collective excitation coupling as suggested by Shen and
Schrieffer
\cite{she} for PES and discussed by DeWilde $et$ $al$ \cite{dew} for STM
and PCT. We have shown earlier that a similar dip structure appears in
the electronic  DOS as a result of the electron-Einstein phonon interaction
\cite{alemaz}. With the established DOS, Eq.(12) we can  quantify
this feature. If we determine $V_{c}\simeq 70 meV$ in the overdoped sample,
Fig.2, and $V_{c}\simeq 80 meV$ in the optimally doped, then the relevant
 polaron
kinetic energy, $E_{c}=eV_{c}-\Delta/2$, appears to be about $39 meV$ and
$42 meV$, respectively. Hence, the emission of any  dispersionless
phonon of  that energy can produce a dip in the polaronic DOS as
described in Ref. \cite{alemaz}. However, $E_{c}$ is also close to the
estimated energy of the magnetic fluctuations thought to be responsible for the
spin-flip neutron scattering at $q=(\pi,\pi)$ ( $41 meV$ peak observed in
$YBa_{2}Cu_{3}O_{7}$ \cite{ros}).  These magnetic fluctuations,
should they be  found in Bi2212, might
contribute to the dip  as well. The polaron-bipolaron
inelastic collisions also
 contribute to the dip in a manner similar to the polaron-phonon
 interaction \cite{alemaz} because $V_{c}$ is close to $\Delta$.
 The polaronic DOS is finite in the
middle of the gap, as discussed above.
 Hence, the polaron with the kinetic energy $E_{c}$
can break a bipolaron into two
single polarons  with the energy about $\Delta/2$ each.

The present theory of tunnelling and PES can be generalised to
describe SIS junctions, $c-axis$ current at high voltage  and the
angle-resolved photoemission (ARPES). In particular, SIS current-voltage
characteristics are obtained by a convolution of the polaronic DOS
with itself. As a result we get two
peaks  around the temperature independent $|eV|=\Delta$,
and the symmetric shape of the SIS conductance, as observed \cite{dew}.
We expect an $S$-shape $I-V$ for the $c-axis$ current $both$ in the
superconducting and normal state as a result of
the bipolaron-breaking at high voltage. The $S$ shape c-axis $I-V$
characteristics were  measured  in Bi2212 and interpreted as
a  `subgap resistance of intrinsic Josephson junctions'
between $superconducting$ $CuO_{2}$
planes\cite{jur}. We believe, that while the notion of `subgap'
 is perfectly correct, one cannot
 refer the observed nonlinearity to the Josephson junctions.
 The nonlinearity is still found at temperatures as high
as $140K$ (Fig.4  of Ref.\cite{jur}) where the $CuO_{2}$ planes are
 definitely in the
normal state, as expected in the framework of the bipolaron theory.
Therefore, there is  no doubt
  that the true normal state $c-axis$ resistivity and the
  upper critical field $H_{c2}$ were measured in
Ref.\cite{alezav} by applying an external magnetic field.
ARPES can be
 described with the spectral function
$A(k,E)$, determined numerically from Eqs (7,8). While
such a feature of
ARPES as  the normal state gap is understood
within the present analysis, the $k$ dispersion will be presented
elsewhere.

In summary, we propose a single-particle spectral function for cuprates,
which describes the  spectral features observed in tunnelling
and photoemission, in particular the temperature independent gap and the
anomalous $gap/T_{c}$ ratio, injection/emission asymmetry both in the
magnitude and in the shape,
zero-bias conductance at zero temperature, and the spectral shape inside and
outside the gap region. The dip-hump features and the temperature/doping
dependence of the tunnelling conductance and PES are discussed as well.

The author highly appreciates enlightening discussions with G.S.
Boebinger, A.M. Campbell,
R.A. Doyle,
O. Fisher, W.Y. Liang, Ch.
Renner, S.G. Rubin,
 J.R. Schrieffer, Z.-X. Shen, M. Springford, and V.N. Zavaritsky. We are
 grateful to M.A. Alexandrov for his assistance in computer calculations.
 We thank E.N. Sladkovskaia for her careful reading of the manuscript.

\centerline{{\bf Figure Captures}}

Fig.1 SIN tunnelling into a single polaron state $P$ and into a bipolaron
state $B$.

Fig.2  Theoretical tunnelling conductance (line) compared with  the STM
conductance ($T=4.2K$)
in overdoped $Bi2212$ ($T_{c}=74.3 K$ \cite{ren}).

Fig.3 Theoretical tunnelling conductance (line) compared with the PCT
conductance ($T=4.2K)$
in optimally doped $Bi2212$ ($T_{c}=95 K$ \cite{dew}). Inset shows
the temperature dependence of the conductance calculated with
 a constant $\epsilon_{0}$ in the gap region.

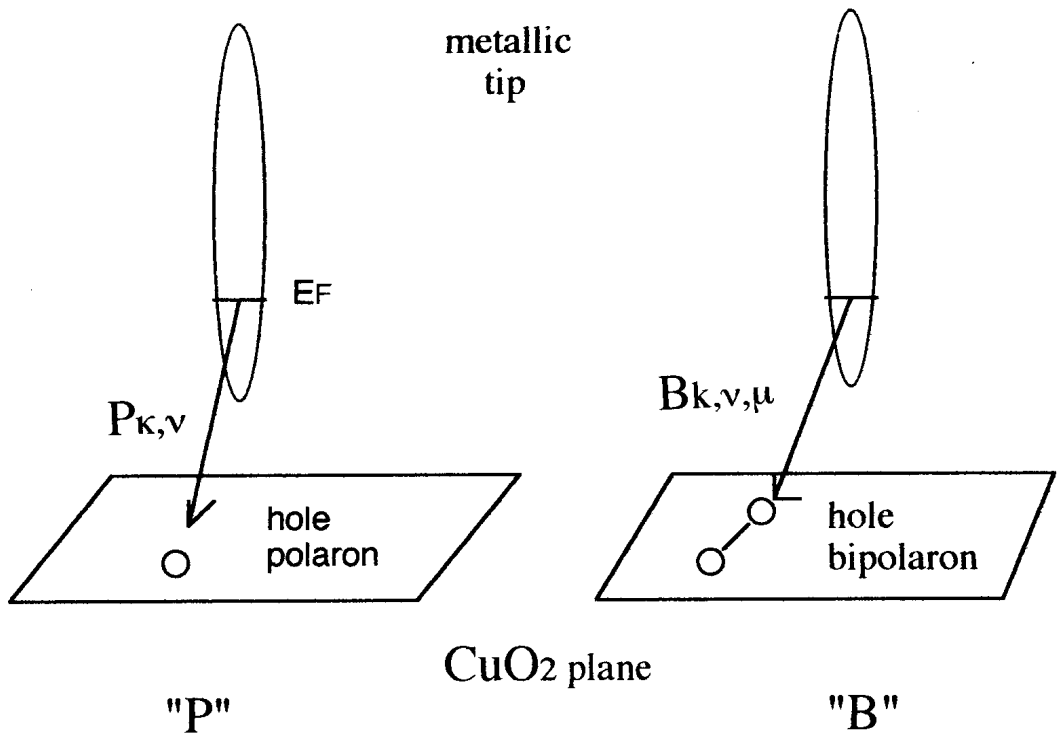
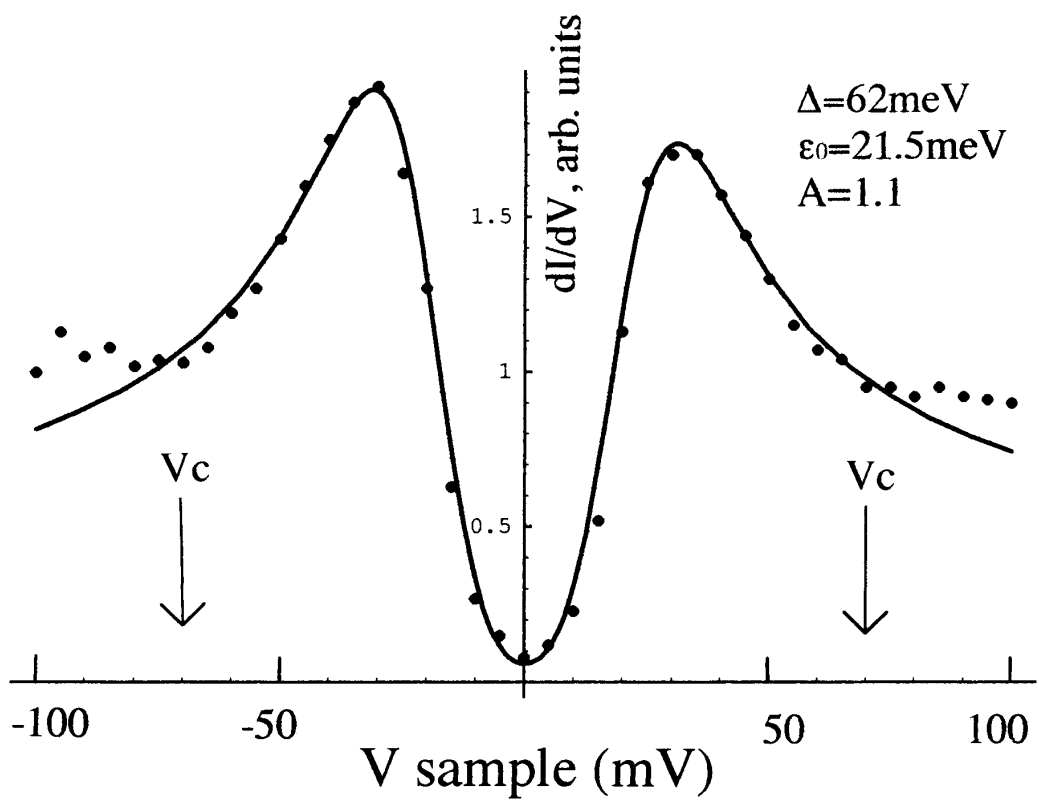
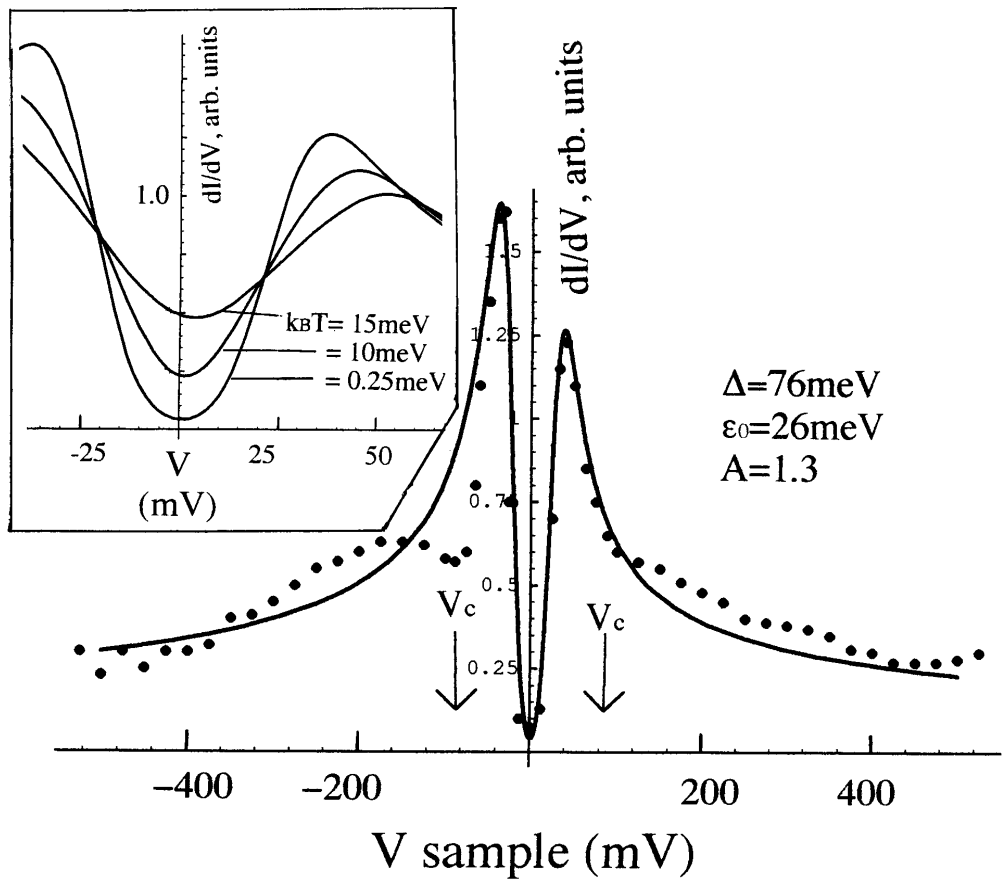
\end{document}